# SweepFinder2: Increased sensitivity, robustness, and flexibility

Michael DeGiorgio[1,*], Christian D. Huber[2], Melissa J. Hubisz[3], Ines Hellmann[4], and Rasmus Nielsen[5]

[1]Department of Biology, Pennsylvania State University, University Park, PA, USA
[2]Department of Ecology and Evolutionary Biology, University of California, Los Angeles, USA
[3]Department of Biological Statistics and Computational Biology, Cornell University, Ithaca, NY, USA
[4]Department Biologie II, Ludwig-Maximilians-Universität München, Planegg-Martinsried, Germany
[5]Department of Integrative Biology, University of California, Berkeley, CA, USA

## Abstract

**Summary:** SweepFinder is a popular program that implements a powerful likelihood-based method for detecting recent positive selection, or selective sweeps. Here, we present SweepFinder2, an extension of SweepFinder with increased sensitivity and robustness to the confounding effects of mutation rate variation and background selection, as well as increased flexibility that enables the user to examine genomic regions in greater detail and to specify a fixed distance between test sites. Moreover, SweepFinder2 enables the use of invariant sites for sweep detection, increasing both its power and precision relative to SweepFinder.

**Availability and implementation:** SweepFinder2 is a freely-available (www.personal.psu.edu/mxd60/sf2.html) software package that is written in C and can be run from a Unix command line.

**Contact:** mxd60@psu.edu

## Introduction

Polymorphism frequency spectra provide sensitive statistics for identifying signatures of positive selection. SweepFinder (Nielsen *et al.* 2005) is a widely used program (Williamson *et al.* 2007, Svetec *et al.* 2009, Pavlidis *et al.* 2010, Li *et al.* 2011) that uses an empirical background frequency spectrum for identifying genomic sites affected by recent positive selection. Specifically, SweepFinder performs a composite likelihood ratio test for positive selection (Kim and Stephan 2002), in which the likelihood of the null hypothesis is calculated from the neutral (or genome-wide) frequency spectrum, and the likelihood of the alternative hypothesis is calculated from a model in which the neutral spectrum was altered by a recent selective sweep.

Footprints of positive selection can be confounded by other evolutionary forces. One important confounding factor that is rarely considered in studies of positive selection is background selection, which is a loss of neutral variation due to purging of linked deleterious alleles by negative selection (Charlesworth *et al.* 1993, Hudson and Kaplan 1995a, Charlesworth 2012). Recent studies have shown that background selection is ubiquitous in humans (McVicker *et al.* 2009, Lohmueller *et al.* 2011, Wilson Sayres *et al.* 2014), with estimates of mean reductions in genetic diversity due to background selection ranging from 19-26% and 12-40% on autosomes and the X chromosome, respectively (McVicker *et al.* 2009). Thus, the influence of background selection on genetic diversity has important ramifications for making inferences about past adaptive processes from patterns of diversity. In particular, when a beneficial allele is carried to fixation by positive selection, there is a substantial decrease in diversity locally in the genome and a reduction in diversity relative to divergence with other species, both of which can span megabases in length (Maynard Smith and Haigh 1974). Background selection can similarly affect diversity levels (Charlesworth *et al.* 1993, Charlesworth *et al.* 1995, Hudson and Kaplan 1995a,b, Nordborg *et al.* 1996, McVean and Charlesworth 2000, Boyko *et al.* 2008, Akashi *et al.* 2012, Charlesworth 2012), particularly in regions of low recombination.

Because patterns of background selection can mimic those of positive selection, methods for identifying signatures of positive selection that are based on diversity reduction alone may be confounded by strong signals of background selection. These conflicting signals have likely contributed to a current debate of the role of recent positive selection in shaping the landscape of human genetic variation (Hawks *et al.* 2007, Williamson *et al.* 2007, Akey 2009, Hernandez *et al.* 2011, Lohmueller *et al.* 2011, Granka *et al.* 2012, Enard *et al.* 2014), emphasizing the need for methods that can identify sweeps while accounting for background selection. Further, because the effects of background selection may be pronounced in regions of low recombination, it is important that methods jointly account for background selection and local recombination rate, which is also expected to affect patterns of a selective sweep.

## SweepFinder2

SweepFinder2, which is based on the statistical framework of SweepFinder (Nielsen *et al.* 2005), jointly accounts for background selection and local recombination rate by modeling the effect of background selection on genetic diversity. It does this by modifying the neutral derived frequency spectrum with respect to *B*-values and by including invariant sites (specifically substitutions), as introduced by Huber *et al*. (2015). *B*-values range from 0 to 1 and are proportional to local reductions in genetic diversity or effective population size due to background selection. McVicker *et al.* (2009) provide a method for inferring *B*-values using comparative data, thereby providing an opportunity for separating background selection from the effect of selective sweeps inferred from within-population polymorphism data. Because background selection reduces diversity by a factor *B*, we multiply each polymorphic frequency class (*i.e.*, allele counts 1,2,…,*n*-1 in a sample of *n*) by *B*, as shown in Figure 1A (Huber *et al*. 2015). Furthermore, because background selection affects diversity relative to divergence with another species, we scale the fixed difference class (*i.e.*, allele count *n*), and then renormalize the frequency spectrum to sum to 1 (Fig. 1A). Note that this effect depends on the current and ancestral population sizes, as well as on the divergence time in generations between the pair of species. Figure 1B illustrates how this procedure modifies the neutral frequency spectrum, such that diversity decreases and the proportion of fixed differences increases with increasing effect of background selection (*i.e.*, decreasing *B*-value).

Our method detects selective sweeps in regions under background selection by scaling the neutral frequency spectrum locally in the genome by estimated *B*-values (Fig. 1), using the scaled spectrum in the null hypothesis, and the spectrum under a model of a selective sweep (accounting for local recombination rate) in the alternative hypothesis (Huber *et al*. 2015). Regions with reductions in diversity and low *B*-values show little evidence of selective sweeps under this test because frequency spectra under the null and alternative hypotheses are similar (Fig. 1C). However, regions with reductions in diversity and relatively high *B*-values may provide evidence of recent selective sweeps, because frequency spectra under the alternative hypothesis will exhibit lower diversity than those under the null hypothesis. In addition, recent positively-selected alleles within regions undergoing background selection can still be detected (Fig. 1D). Furthermore, changes in *B*-values across the genome can be incorporated by modifying frequency spectra to preserve the spatial structure in genetic variation leveraged by SWEEPFINDER.

SWEEPFINDER2 is the first method that accounts for the effects of negative selection on diversity when searching for adaptive alleles. Additionally, it incorporates novel features that provide the user with increased flexibility. Thus, our new composite likelihood ratio test generalizes the one implemented in SWEEPFINDER (Nielsen *et al*. 2005), and provides a substantial improvement in power and flexibility to the popular SWEEPFINDER software.

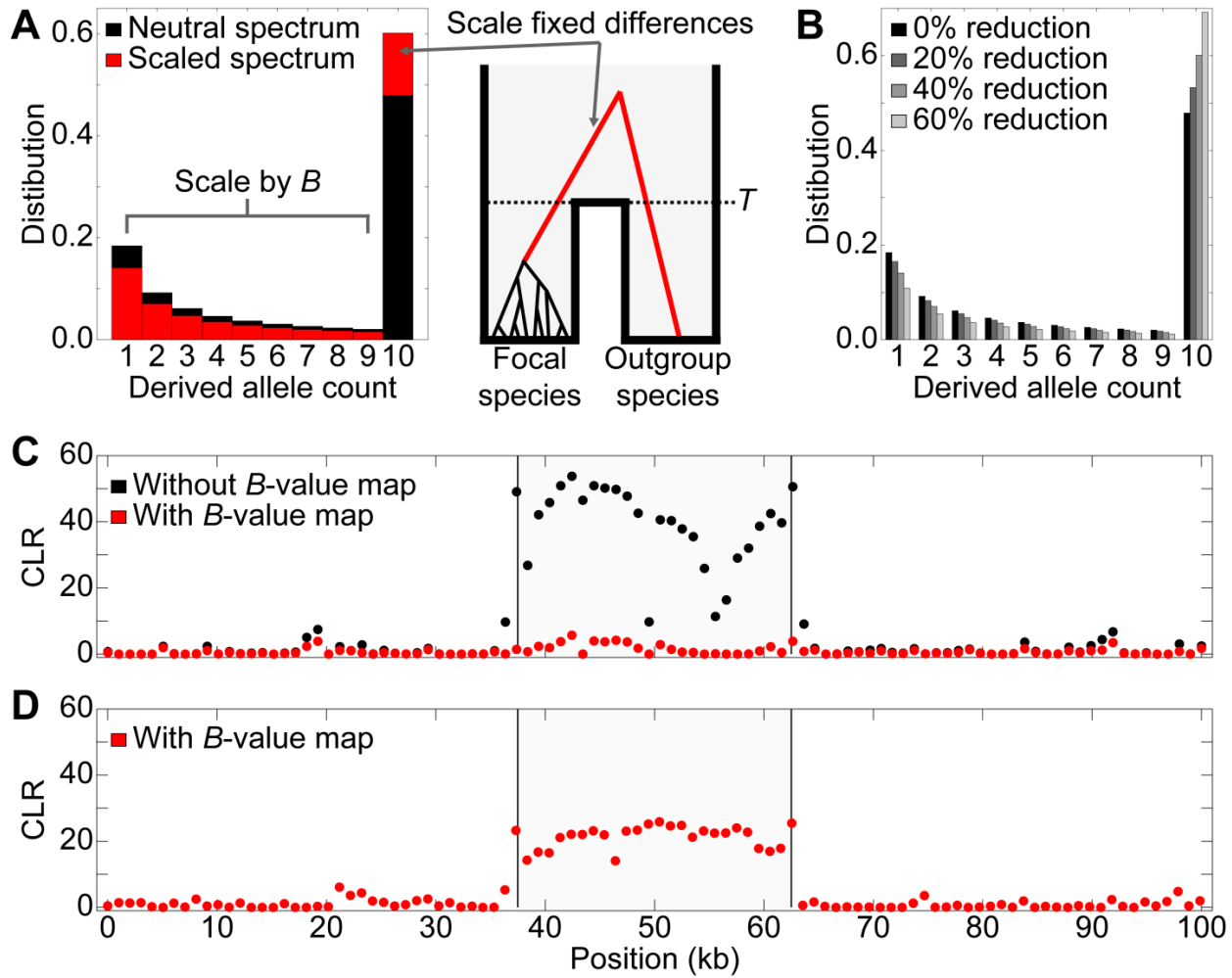

**Fig. 1.** Generating derived frequency spectra from a neutral frequency spectrum under background selection in a sample of 10 alleles and an outgroup sequence. (A) Polymorphic sites (allele counts 1-9) are scaled by a factor $B$, reducing diversity by 1-$B$. The proportion of fixed sites (allele count 10) is scaled by $(T+2BN/n)/(T+2N/n)$, and the spectrum is then normalized to sum to 1. The scaling factor for the fixed difference class assumes a model in which a pair of species split $T$ generations ago, with all populations having effective size $N$ (SWEEPFINDER2 implementation permits unequal sizes). (B) Modified frequency spectra for 0, 20, 40, and 60% reductions in diversity due to background selection ($B$-values of 1.0, 0.8, 0.6, and 0.4, respectively). As $B$-value decreases, the level of diversity decreases, and the ratio of diversity to divergence decreases. (C, D) Composite likelihood ratio test statistics as a function of position along a sequence without (C) and with (D) a fixed selective sweep in the center of the sequence. The gray region represents a reduction in recombination rate by two orders of magnitude. Including the $B$-value map decreases false inferences of positive selection (C), yet still can identify positively-selected alleles in regions with background selection (D).